\documentclass[preprint]{aastex}

\usepackage{color}

\shorttitle{Granulation and faculae on Sun-like stars}
\shortauthors{Karoff et al.}

\begin{document}
\title{Observations of intensity fluctuations attributed to granulation and faculae on Sun-like stars from the {\it Kepler} mission}
\author{C. Karoff}
\affil{Stellar Astrophysics Centre, Department of Physics and Astronomy, Aarhus University, Ny Munkegade 120, DK-8000 Aarhus C, Denmark}
\email{karoff@phys.au.dk}
\author{T.~L. Campante}
\affil{Centro de Astrof\'isica and Faculdade de Ci\^encias, Universidade do Porto, Rua das Estrelas, 4150-762 Porto, Portugal}
\affil{School of Physics and Astronomy, University of Birmingham, Edgbaston, Birmingham, B15 2TT, UK}
\author{J. Ballot}
\affil{CNRS, Institut de Recherche en Astrophysique et Plan\'etologie, 14 avenue \'Edouard Belin, 31400, Toulouse, France}
\affil{Universit\'e de Toulouse, UPS-OMP, IRAP, Toulouse, France}
\author{T. Kallinger}
\affil{Instituut voor Sterrenkunde, K.U. Leuven, Celestijnenlaan 200D, 3001 Leuven, Belgium}
\affil{Institute for Astronomy, University of Vienna, T\"urkenschanzstrasse 17, 1180 Vienna, Austria}
\author{M. Gruberbauer}
\affil{Institute for Computational Astrophysics, Department of Astronomy and Physics, Saint MaryÕs University, B3H 3C3 Halifax, Canada}
\author{R.~A. Garc{\'{\i}}a}
\affil{Laboratoire AIM, CEA/DSM-CNRS-UniversitŽ Paris Diderot; IRFU/SAp, Centre de Saclay, 91191 Gif-sur-Yvette Cedex, France}
\author{D.~A. Caldwell \& J.~L.~Christiansen}
\affil{SETI Institute/NASA Ames Research Center, Moffett Field, CA 94035}
\and
\author{K. Kinemuchi}
\affil{Bay Area Environmental Research Inst./NASA Ames Research Center, Moffett Field, CA 94035}

\begin{abstract}
Sun-like stars show intensity fluctuations on a number of time scales due to various physical phenomena on their surfaces. These phenomena can convincingly be studied in the frequency spectra of these stars -- while the strongest signatures usually originate from spots, granulation and p-mode oscillations, it has also been suggested that the frequency spectrum of the Sun contains a signature of faculae. 

We have analyzed three stars observed for 13 months in short cadence (58.84 seconds sampling) by the $Kepler$ mission. The frequency spectra of all three stars, as for the Sun, contain signatures that we can attribute to granulation, faculae, and p-mode oscillations. 

The temporal variability of the signatures attributed to granulation, faculae and p-mode oscillations were analyzed and the analysis indicates a periodic variability in the granulation and faculae signatures -- comparable to what is seen in the Sun. 
\end{abstract}


\section{Introduction}
Low-mass main sequence stars (Sun-like stars) have an outer convection zone, which means that they show granulation on their surface. The outer convection zone is responsible for exciting acoustic oscillations inside these stars, which, at least in the stars with masses close to the Sun, have observable amplitudes at the surfaces of these stars \citep{1999A&A...351..582H}. When observing the disk-intergrated intensity of these stars as a function of time, the resulting light curves will thus include the signatures of both granulation and oscillations. These signatures are usually hard to identify directly in the light curves as granulation, which has the largest amplitude, is a strongly non-coherent phenomenon. The same applies to oscillations in Sun-like stars as their oscillations are damped, which means that they are also not coherent on long timescales \citep{1994ApJ...424..466G}. An exception from this are evolved stars, which can have clear signatures of both granulation and oscillations in their light curves \citep{1975ApJ...195..137S}.

Granulation and oscillations are therefore usually studied in Sun-like stars by analyzing the frequency spectrum of their light curves. These frequency spectra show a characteristic decline in power with increasing frequency caused by granulation. The location in frequency of this characteristic decline gives the characteristic time scale of granulation -- equivalent to the life time or the turn-over time of the granules, which has been measured in the Sun \citep{1985shpp.rept..199H}, sun-like stars \citep{2008Sci...322..558M} and evolved red-giant stars \citep{2011ApJ...741..119M}.

The characteristic frequency of the p-mode oscillations in Sun-like (and evolved) stars (i.e. the frequency of maximum power) is proportional to the atmospheric acoustic cut-off frequency which is given as the sound speed divided by twice the pressure scale height near the surface -- assuming the atmosphere is isothermal \citep{2011A&A...530A.142B}. Traditionally, it has been assumed that the characteristic time scale of granulation would also scale with the atmospheric acoustic cut-off frequency \citep{1991ApJ...368..599B, 1995A&A...293...87K, 2011A&A...529L...8K}. This means that there would be a linear relation between the atmospheric acoustic cut-off frequency and the difference between the characteristic time scale of granulation and the p-mode oscillations.

It has been known for some time that the frequency spectrum of the Sun shows another feature in the region located between the characteristic decline in power caused by granulation and the bump caused by the p-mode oscillations \citep{1993ASPC...42..111H}. Lately, it has been proposed that this feature is likely caused by bright faculae on the surface of the Sun \citep{2012MNRAS.tmp.2467K}. 

At low frequencies, peaks can be seen caused by the rotational modulation of long-lived sunspots \citep{2004A&A...425..707L}. At frequencies below 1 $\mu$Hz a linear decline in power with increasing frequency can be seen, which is caused by the weak coherency of activity-related phenomena such as sunspots \citep{1993ASPC...42..111H}. At frequencies higher than the atmospheric acoustic cut-off frequency evidence is seen for the so called high-frequency peaks, which could be traveling waves or chromospheric oscillations \citep{1998ApJ...504L..51G, 2005ApJ...623.1215J, 2007MNRAS.381.1001K, 2008ApJ...678L..73K, 2009ASPC..416..233K}.

At least two different interpretations of the physical processes responsible for the different features in the solar frequency spectrum can be found in the literature. In the first interpretation, which has been used by \citet{1998ESASP.418...83A}, \citet{1998ESASP.418...91A}, \citet{2004ESASP.538..215A} and  \citet{2009A&A...495..979M}, the different features are explained by: rotation, supergranulation, mesogranulation, granulation and oscillations (see Table~1). In the second interpretation, which has been used by \citet{1993ASPC...42..111H}, \citet{2002ESASP.506..897V} and \citet{2012MNRAS.tmp.2467K}, they are explained by: rotation, granulation, bright points (or faculae) and oscillations (see Table~2). Though this is a rough separation of all these studies into only two different interpretations, it exemplifies the general difference between these studies -- that the studies using the first interpretation ascribe the feature just below 1000 $\mu$Hz to mesogranulation and the feature just above to granulation, whereas the studies using the second interpretation ascribe the feature just below 1000 $\mu$Hz to granulation and the feature just above to bright points or faculae.

Following \citet{1998A&A...330.1136R} the three different convective scales can be identified as: Supergranulation characterized by a mean size of 20 to 50 Mm, mesogranulation by a mean size of 5 to 10 Mm and granulation by a mean size of around 1000 km. The coherence time-scale of granulation is between 2 and 6 min. \citep[][and references herein]{2004A&A...428.1007D}. This agrees with the interpretation by \citet{1993ASPC...42..111H}, \citet{2002ESASP.506..897V}, and \citet{2012MNRAS.tmp.2467K} that the signature just below 1000 $\mu$Hz is due to granulation. Mesogranulation and supergranulation would have to have significant longer coherence time-scales and thus manifest themselves at frequencies significantly below 1000 $\mu$Hz. No such manifestations seem to be visible in the observed frequency spectrum of the Sun \citep{2012MNRAS.tmp.2467K}.

That the manifestation of mesogranulation and supergranulation is not observable in the frequency spectrum of the Sun does not mean that mesogranulation and supergranulation does not contribute to the background in the frequency spectrum. The contribution is just not large enough to be observable. The same is true for the second population of granules with coherence time-scales of around 1 min. identified by \citet{2004A&A...428.1007D}. It has been shown by \citet{2012MNRAS.tmp.2467K} that this second population of granules, which would be both more numerous and with a lower brightness contrasts, would not be observable in the frequency spectrum, as the amplitude of the signature in the frequency spectrum scales with the brightness contrast over the square root of the number of granules. We thus follow the second interpretation in this paper.

Here we analyze observations from the $Kepler$ spacecraft of the three Sun-like stars:  KIC 6603624, KIC 6933899 and KIC 11244118. KIC 6603624 was one of the three stars presented in the first results on solar-like stars based on $Kepler$ observations \citep{2010ApJ...713L.169C} and already in that work it was noted that the frequency spectrum of this star shows a number of different components. One component possible originating from bright faculae (as seen in the Sun). The eigenmode frequencies of all three stars were analyzed in detail by \citet{2012arXiv1202.2844M} and the resulting stellar parameters are shown in Table~3. Note that the uncertainties in Table~3 are formal uncertainties only, which do not include estimates of systematic contributions. 

The three stars were selected partly because they were among the 22 stars modelled by \citet{2012arXiv1202.2844M}, which provides us with precise stellar parameters from asteroseismology and partly because they all show frequency spectra with features close to those observed in the Sun. The three stars are not the only Sun-like stars observed by {\it Kepler} that show the signature that we attribute to faculae in their spectra. 

The outline of this paper is the following: In section 2 we describe the 13 months of observations from the $Kepler$ spacecraft that were used in this study and how these observations were pre-processed. A description of the analysis of the frequency spectra is given in section 3, including how we measure the parameters of the components that we attribute to granulation, faculae and p-mode oscillations. The results are presented in section 4 and conclusion are given in section 5.

\section{Observations}
The three stars were all observed in short cadence  \citep[58.85 sec. sampling,][]{2010ApJ...713L.160G} by the $Kepler$ mission \citep{2010ApJ...713L..79K}. The specific observing times are given in Table~4. The light curves were prepared as described in \citet{2010ApJ...713L..87J}  and \citet{2011MNRAS.414L...6G}, which removes most instrumental artefacts. The one artefact that was left in the light curves was a general declining trend. As all light curves are normalized to one, it is thus problematic to connect different quarters of observations. This is partly why we have chosen to calculate individual frequency spectra of all months and then combining these spectra to an average spectrum -- and because we do not need high-frequency resolution for studying granulation, faculae and the envelope of the p-mode oscillations.

Apart from the declining trend, all three stars show RMS scatter with an amplitude of around one hundred ppm in the time domain and the noise level is relatively constant from month to month. The first month in quarter 2 contains a gap, as does the last month of quarter 7 and the second month of quarter 8. Nothing is done to these gaps in the analysis as the normalization to power density (see below) accounts for the missing observations. This means that the measured parameters are not affected by the gaps.

\section{Analysis}
The analysis of the signatures attributed to granulation, faculae, and p-mode oscillations in the frequency spectra consists of three steps. First, the frequency spectra are calculated from the light curves, then secondly, the resulting observed spectra are compared to a model in order to find the best model and thirdly, the parameters of the components attributed to granulation, faculae and p-mode oscillation are measured. 

The power density spectra were calculated using the least-squares method \citep{1976Ap&SS..39..447L, karoff2008}. Each spectrum was normalized by the effective observation length given as the reciprocal of the area under the window function in order to convert the spectra into power density and insure that the spectra obey Parseval's theorem. 

We have only analyzed the spectra in the frequency range between 100 $\mu$Hz and up to the frequency of maximum power \citep[as in][]{2012MNRAS.tmp.2467K}. The reason for the lower limit is that this part of the spectrum is often affected by instrumental and rotational effects, which are not the subject of this study. The reason for the higher limit is discussed in details by \citet{2012MNRAS.tmp.2467K} and is related to the choice of model, as we discuss below.

\subsection{The model}
Traditionally, the signatures attributed to granulation and faculae have been evaluated using the model suggested by \citet{1985shpp.rept..199H}:
\begin{equation}
f(\nu)=\frac{4\sigma^2\tau}{1+(2\pi\nu\tau)^2},
\end{equation}
where $f(\nu)$ is the power density at frequency $\nu$, $\sigma$ is the amplitude of the signature and $\tau$ is the characteristic time scale. 

As discussed in \citet{2012MNRAS.tmp.2467K} various corrections have been made to this equation in order to be able to model the frequency spectrum of the Sun. Here we adopt the model from \citet{2012MNRAS.tmp.2467K}:
 \begin{eqnarray}
 f(\nu)&=&\frac{\zeta_{\rm gran} \sigma_{\rm gran}^2\tau_{\rm gran}}{1+(2\pi\nu\tau_{\rm gran})^{3.5}}+\frac{\zeta_{\rm fac} \sigma_{\rm fac}^2\tau_{\rm fac}}{1+(2\pi\nu\tau_{\rm fac})^{6.2}} \nonumber \\
 &&+H_{\rm osc}{\rm exp}\left[\frac{(\nu-\nu_{\rm max})^2}{2w^2}\right]+Noise.
 \end{eqnarray}
We have included the normalization constants $\zeta_{\rm gran}$ and $\zeta_{\rm fac}$, which we will discuss in the next section. $H_{\rm osc}$ is the height of the p-mode oscillation envelope, $\nu_{\rm max}$ is the frequency of maximum power and $w$ is the width of the p-mode oscillation envelope. A white-noise term has also been added to the model.

For the three stars in this study we adopt the solar value of the exponents in eq.~2 -- i.e. $-3.5\pm0.3$ for granulation and $-6.2\pm0.7$ for faculae \citep{2012MNRAS.tmp.2467K}. The size of the exponents is a measurement of the amount of memory in the physical process responsible for the component. A larger exponent means less memory in the process. As the physical processes responsible for granulation and faculae are expected to be the same on other Sun-like stars, with global parameters close to the solar ones, the two exponents are expected to be close to solar values. 

As discussed by \citet{2012MNRAS.tmp.2467K} the model in eq.~1 fails to reproduce the observed solar acoustic background for frequencies higher than the atmospheric acoustic cut-off frequency. The reason for this is that granulation cannot be modelled with turbulent cascades \citep{1997A&A...328..229N} as it is done in the drift model by \citet{1985shpp.rept..199H}. Turbulence shows a distribution with a slope of around $-2$ in power, convection on the other hand has a lower limit in the time domain on which changes can take place. This means that on small time-scales (or at high frequency) convection is not noisy whereas turbulence is. Therefore in order to model the observed solar acoustic background for frequencies higher than the atmospheric acoustic cut-off frequency an extra term has to be included in eq.~1 (or 2) as it was done in \citet{karoff2008}. In order not to have to make assumptions about the behavior of this extra term in other stars, we have chosen not to model the high frequency part of the spectra and this is why we only analyze the spectra up to the frequency of maximum power.

\subsection{Normalization}
The amplitudes of the components attributed to granulation and faculae in eq.~2 can be understood as the variance these signals have in the time series. This is insured by Parseval's theorem, but in order for eq.~2 to obey Parseval's theorem, eq.~2 needs to be normalized using the normalization constants $\zeta_{\rm gran}$ and $\zeta_{\rm fac}$.
 
Parseval's theorem says that:
\begin{equation}
\sum_{\nu=0}^{\nu=\nu_{Ny}^{}}f(\nu)\delta \nu=\sum_{t=t_0}^{t=t_N}\frac{|o(t)-\bar{o}|^2}{N}\delta t=\sigma^2,
\end{equation}
where $\nu_{Ny}$ is the Nyquist frequency, $N$ the number of observations, $o(t)$ is the observed intensity at time $t$ and $\bar{o}$ the mean value of the observations. $\delta t$ is the time step and $\delta \nu$ is the frequency resolution ($\delta \nu = \frac{1}{N \delta t}$). The first sum should be summed over all natural frequencies, whereas the second sum should be summed over all time steps. As we assume the different contributions to the frequency spectra -- granulation, faculae and the p-mode oscillation envelope, to be non-coherent and uncorrelated, Parseval's theorem can also be written as a sum of $k$ different contributions:
\begin{equation}
\sum_{\nu=0}^{\nu=\nu_{Ny}^{}}\sum_k{f_k^{}(\nu)} \delta \nu = \sigma^2 = \sum_k{\sigma_k^2}
\end{equation}
and the different contributions can thus be normalized individually to Parseval's theorem, as they are additive in power:
\begin{equation}
\sum_{\nu=0}^{\nu=\nu_{Ny}^{}}f_k^{}(\nu)\delta \nu =  \sigma_k^2
\end{equation}
Following \citet{2009A&A...495..979M} it is thus possible to calculate the normalization constants $\zeta_{\rm gran}$ and $\zeta_{\rm fac}$ from eq. 2:
 \begin{equation}
 \sum_{\nu=0}^{\nu=\nu_{Ny}^{}}\left[ \frac{\zeta \sigma_k^2\tau_k^{}}{1+(2\pi\nu\tau_k^{})^{\alpha_k}} \right]\delta \nu =  \sigma_k^2,
  \end{equation} 
 where $\alpha_k$ are the exponents in eq.~2 -- i.e. 3.47 for granulation and 6.20 for faculae. Assuming that $\nu_{Ny}^{}\tau \gg 1$ it can be shown that:
  \begin{equation}
\zeta_k = 2\alpha_k {\rm sin} \left(\frac{\pi}{\alpha_k} \right)
 \end{equation} 
 Using this formulation we obtain: $\zeta_{\rm gran} = 5.46$ and $\zeta_{\rm fac} = 6.02$.

\subsection{Minimization}
The model described above was matched to the observed frequency spectra using maximum-likelihood estimators as described in \citet{2012MNRAS.tmp.2467K} by calculating the logarithmic likelihood function $\ell$ between $N$ independent measurements $x_i$ -- i.e. power density at a given frequency, and the model $f_i$ given by a set of parameters ${\bf \lambda}$:
\begin{equation}
\ell=-\sum_{i=1}^N{{\rm ln} \ p(x_i,{\bf \lambda})},
\end{equation}
where $p(x_i,{\bf \lambda})$ is the probability density function, which is obtained from \citet{2004A&A...428.1039A}:
\begin{equation}
p\left[\mathcal{S}_i(x,n),f_i({\bf \lambda})\right]=\frac{\mu_i^{\nu_i}}{\Gamma(\nu_i)}\mathcal{S}_i(x,n)^{\nu_i-1}e^{-\mu \mathcal{S}_i(x,n)},
\end{equation}
where ${S}_i(x,n)$ is the observed spectrum binned over $n$ bins, $\Gamma$ is the Gamma function and $\mu_i$ and $\nu_i$ are given as:
\begin{equation}
\mu_i=\frac{(n+1)\displaystyle\sum\limits_{k=i-n/2}^{k=i+n/2}{f_k({\bf \lambda})}}{\displaystyle\sum\limits_{k=i-n/2}^{k=i+n/2}{f_k^2({\bf \lambda})}},
\end{equation}
and
\begin{equation}
\nu_i=\frac{\left[\displaystyle\sum\limits_{k=i-n/2}^{k=i+n/2}{f_k({\bf \lambda})}\right]^2}{\displaystyle\sum\limits_{k=i-n/2}^{k=i+n/2}{f_k^2({\bf \lambda})}}.
\end{equation}
Note that $\nu_i$ is different from the $\nu$ used for the frequencies in the frequency spectra.

The general idea is here to compare a binned version of the observed spectra ${S}_i(x,n)$ to the model in order to calculate the likelihood -- instead of assuming that the differences between the observed frequency spectra and the model are given either by a normal distribution or a $\chi^2$ distribution with 2 degrees of freedom. The spectrum should be binned over so many bins that the model becomes a good representation of the observed spectrum. On the other hand $n$ should not be so large that the different features in the frequency spectra cannot be observed. We have thus used $n=100$. If $n$ is close to one eq. 9 reduces to the well known case from the modeling of individual p-mode oscillation modes where the differences between the observed frequency spectra and the model are given by a $\chi^2$ distribution with 2 degrees of freedom \citep{1994A&A...287..685G}.

The minimization of the likelihood function was done using the simplex method as applied in the AMOEBA function \citep{1992nrfa.book.....P}.

\subsection{Uncertainties}
Formal uncertainties can be calculated as the diagonal elements of the inverse of the Hessian matrix, but such uncertainties are only internal uncertainties that says something about the curvature of the minimized likelihood function. 

The uncertainties also have to take into account the averaging that is performed in the modeling. First we average individual spectra to an average spectrum and then we bin this (and the individual spectra) over $n$ bins. While the first effect can be accounted for by dividing the uncertainties by the square root of the number of individual spectra \citep{2003A&A...412..903A} it is not straight forward how to account for the second effect. 

If the differences between an observed frequency spectrum sampled at the natural frequencies and the model were assumed to be given by a $\chi^2$ distribution with 2 degrees of freedom and this spectrum was binned over $n$ bins, then we know that the differences between the binned spectrum and the model would be given by a $\chi^2$ distribution with 2$n$ degrees of freedom. In that case the uncertainties would have to be adjusted accordingly \citep{2012arXiv1204.3147A}.  As the probability density function that we have used (eq. 9) accounts for the binning, it is, on the other hand, not clear if any corrections are need to the uncertainties.

To avoid inconsistencies in the quoted uncertainties, we have chosen to give the average values from the results of different monthly, individual spectra rather than results of an average spectrum. In this way we can also give the uncertainties as the uncertainties on the mean value. This is on the other hand not possible to do on the temporal results, so here the error bars only represent formal uncertainties.    

\section{Results}
Using the method described above we have modelled a summed spectrum calculated as the sum of individual spectra for each of the 13 months of observations for all three stars. These summed spectra were used to calculate the significance of the component attributed to faculae using the formulation described above. The average spectra are shown in Fig.~1 together with the model with (red line) and without (blue line) the component attributed to faculae. We have also analysed the individual spectra for each of the 13 months of observations for all three stars and the measured mean values and their temporal variability are shown in Figs.~4--7.

\subsection{Significance of the component attributed to faculae}
The significance of the signature of the component attributed to faculae was calculated using the logarithmic likelihood ratio $\Lambda$:
\begin{equation}
{\rm ln} \ \Lambda = \ell(\lambda_{p+q})-\ell(\lambda_{p}),
\end{equation}
where $p$ is the number of free parameters in the model without the component attributed to faculae (i.e. 6) and $q$ is the number of additional parameters in the model with the component attributed to faculae (i.e. 2), and comparing the value of $-2$ln$\Lambda$ to a $\chi^2$ distribution with $q$ degrees of freedom as in \citet{2012MNRAS.tmp.2467K}.

A visual inspection of the frequency spectra reveals that the model with the component attributed to faculae matches the observed spectra better than the model without such a component in all three cases. This is also reflected in the returned logarithmic likelihood ratios, which are shown in Table~5. The low ratios mean that the significance of the component attributed to facular is one in all three stars  -- i.e. the probability of having a logarithmic likelihood ratio smaller than e.g. -12 returned from a $\chi^2$ distribution with 2 degrees of freedom is larger than 0.99999. 

In order to test that the measured significances of the faculae components were not affected by using solar values for the exponents in eq.~2 we also calculated the logarithmic likelihood ratios allowing the exponents to vary around the solar values. This did change the returned logarithmic likelihood ratios, which are also shown in Table~5, but not the conclusion that the significance of the component attributed to faculae is one in all three stars. In other words it is clear that the conclusion that the component attributed to faculae is significantly present in all three stars is valid whether the exponents are fixed to solar values or allowed to change freely.

When the exponents were fixed to solar values, the exponent of the granulation component were the same in both models, but when the exponents were allowed to change freely this was not the case -- here the values of the granulation exponent were generally lower in the model with no facular component. Generally, for all the tests we did, the exponents were returned with values between 3 and 5 for the component attributed to granulation and 6 and 8 for the component attributed to faculae. For the model without the faculae component the exponents were returned with values between 1.7 and 4.

The returned logarithmic likelihood ratios also agrees with the impression from a visual inspection -- that the signature of the component attributed to faculae is strongest in KIC 6603624 and weakest in KIC 6933899. In fact, as can be seen in Fig.~1, it was not possible to get a satisfactory agreement between the model with no component attributed to faculae and the observed spectrum of KIC 6603624.

\subsection{Mean values of the measured parameters}
The mean values of the measured parameters and their uncertainties were calculated from the values measured in the 13 spectra of the individual months (see Table~6).

It is well known that there exist a relation between the frequency of maximum power and the large frequency separation, which again depends on the mean stellar density \citep{2009MNRAS.400L..80S}. This agrees nicely with the fact that KIC 6603624 with a density almost twice as large as the two other stars is also the stars with the largest frequency of maximum power (see Table~6). The mean density of KIC 6603624 is still lower than the mean density of the Sun and so is the frequency of maximum power.

Even though three stars is far from enough for calculating scaling relations we have tried to compare both the time scales and amplitudes of the various components to the frequency of maximum power in Figs.~2 \& 3. It is seen that the time scales of both the components scale inversely with the frequency of maximum power (Fig.~2). We do not see any clear relation between the amplitudes of the components and the frequency of maximum power (Fig.~3). This contradicts the predictions by \citet{2011ApJ...732...54C}: that the amplitudes should scale inversely with the square of the frequency of maximum power. The reason for this is most likely that we only analyze three stars in this study. 

\subsection{Temporal variability}
We have measured the temporal variability of the background and oscillation envelope parameters in the 13 spectra of the individual months for each star. The temporal variability of the parameters of the components attributed to granulation and faculae are shown in Figs.~4--7. The white noise component is shown for comparison in Fig.~8.

In order to test if any of the parameters show periodic variability we found the highest peak in the periodogram of the parameters as a function of time and performed a simple test of the significance of this peak. The test was performed by measuring the amplitude of the highest peak in one million artificially generated periodograms. These periodograms were generated by taking the uncertainties ascribed to each measurements at each time step (calculated as the diagonal elements of the inverse of the Hessian matrix) and multiply them with a normally-distributed random number with a mean of zero and a standard deviation of one. The significance was then calculated as the fraction of cases where the highest peak in these one million artificial generated periodograms had an amplitude lower than the amplitude of the highest peak in the observed periodogram.

Here we adopt the definition that a period with a significance level above 0.99 is significant, whereas a period with a significance level above 0.95 is marginal significant. This leads to the conclusion that we see a significant period of $322 \pm 11$ days (error bars calculated assuming normal distributed noise on the measured amplitudes) in the amplitude of the component attributed to granulation in KIC 6603624. Marginal significant periods are found in the amplitude of the component attributed to faculae in KIC 6603624 ($257 \pm 11$ days) and in the time scale of the component attributed to granulation in KIC 6933899 ($233 \pm 13$ days). The significant levels of the highest peaks in the periodograms of the temporal variability of the granulation and faculae parameters for the three stars are given in Figs.~4--7. No significant or marginal significant periods were found in the parameters of the envelope of the p-mode oscillations.

The periods of the periodic variability seen in the components attributed to granulation and faculae in KIC 6603624 and KIC 6933899 are between two hundred and three hundred days. This could be equivalent to the quasi-annual modulation of the amplitude of the component attributed to granulation in the frequency spectrum of intensity observations of the Sun with the Variability of solar IRradiance and Gravity Oscillations \citep[VIRGO,][]{1995SoPh..162..101F} instrument on the Solar and Heliospheric Observatory \citep[$SOHO$,][]{2012MNRAS.tmp.2467K}. Unfortunately, it has been impossible to study such a modulation in velocity observations from the with Global Oscillations at Low Frequency \citep[GOLF,][]{1995SoPh..162...61G} instrument also on $SOHO$ due to a variation of the observation height in the solar atmosphere induced by the orbital period of the satellite \citep{2008A&A...490.1143L}.

Even though non of the identified periods are identical we did calculate the linear Pearson correlation coefficients between the granulation and faculae parameters (see Table~7). The numerical values of these correlation coefficients were generally very low. The largest one being between the amplitude of the component attributed to faculae and the time scale of the component attributed to granulation in KIC 6922899 with a value of 0.74. The low levels of these correlation coefficients verify that there are no correlations between the different parameters and thus that the identified periods are intrinsic to the amplitude of the component attributed to granulation in KIC 6603624, the amplitude of the component attributed to faculae in KIC 6603624 and the time scale attributed to granulation in KIC 6933899, respectively.

\subsection{Comparison of different analysis methods}
It was shown by \citet{2011ApJ...741..119M} that different analysis methods with different free parameters, models, data and number of components can provide different values of the estimated parameters, but that, for a given method, the results and the trends are consistent. Though we do not want to redo this analysis here it is still interesting to investigate if the component attributed to faculae can also be found using different analysis methods.

We therefore redid the analysis of the frequency spectra of the three stars using the CAN and A2Z methods from \citet{2011ApJ...741..119M} \citep[see][for a detailed description of the methods]{2010A&A...511A..46M, 2010A&A...522A...1K} and a method that used the same model as described here, but no binning of the spectra and thus assuming that the differences between the observed frequency spectra and the model were given by a $\chi^2$ distribution with 2 degrees of freedom. 

The result of the comparison was that the component attributed to faculae was significantly present in the observed frequency spectra with a significance close to 1 in all three stars in all the tests where convergence to physically meaningful parameters could be reached. 

The comparison {\it also} showed that the uncertainties on the mean parameter values are realistic -- i.e. they are comparable in size to the difference between the different analysis methods and the uncertainties returned by the {\it MultiNest} algorithm used in the CAN methods (which have been shown in brackets for comparison in Table~6). The error bars on the temporal results calculated from the diagonal elements of the inverse of the Hessian matrix are on the other hand generally too low \citep[as also noted by ][]{2011ApJ...741..119M}.

\section{Conclusions}
The analysis of the observed frequency spectra of the three Sun-like stars KIC 6603624, KIC 6933899 and KIC 11244118 from 13 months of high-precision, high-cadence photometric observations from the $Kepler$ spacecraft has revealed signatures of what is likely granulation, faculae, and p-mode oscillations in all three stars. 

It is seen that the characteristic frequencies (or equivalent, time scales) of the components attributed to both granulation and faculae inversely scale with the frequency of maximum power and that the frequency of maximum power scale with the mean density of the stars as expected as the atmospheric acoustic cut-off frequency scales with the large frequency separation \citep{2009MNRAS.400L..80S}.

The analysis of the temporal variability of the measured parameters revealed periodic variability in the amplitude of the component attributed to granulation in KIC 6603624 with a significance level above 0.99 and a period of $322 \pm 11$ days. Marginal significant periodic variability with a significance level above 0.95 were found in the amplitude of the component attributed to faculae in KIC 6603624 with a period of $257 \pm 11$ days and in the time scale of the component attributed to granulation in KIC 6933899 with a period of $233 \pm 13$ days.

The temporal variability in KIC 6603624 and KIC 6933899 could have a origin similar to the periodicities of around 1--2 yrs. seen in a number of indices related to solar activity, including: Sun-spot number and neutrino flux \citep{1979Natur.278..146S}, Galactic cosmic-ray intensities \citep{1996SoPh..167..409V}, flare occurrence \citep{1994AdSpR..14..721A}, solar wind velocities \citep{1994GeoRL..21.1559R}, $aa$ geomagnetic indices \citep{2003SoPh..212..201M}, coronal hole area and radio emission \citep{2008AdSpR..41..297V}, the rotation of the Sun near the base of its convective zone \citep{2000Sci...287.2456H} and lately in the residuals of the p-mode frequency shifts \citep{2010ApJ...718L..19F, 2012MNRAS.420.1405B, 2012arXiv1210.6796S}. 

The periods of the periodic variability in KIC 6603624 (and KIC 6933899) are between two hundred and three hundred days, which is somewhat lower than the 1--2 yrs. periodicities observed in the indices related to solar activity. This agrees with the lower densities of these stars compared to the Sun, but it is not clear if it agrees with the period of any possible dynamo in these stars. 

\acknowledgments
We would like to thank the referee for thoughtful comments, which significantly improved the paper. Funding for this Discovery mission is provided by NASAs Science Mission Directorate. The authors wish to thank the entire {\it Kepler} team, without whom these results would not be possible. CK acknowledged support from the Carlsberg foundation. TLC acknowledges financial support from
project PTDC/CTE-AST/098754/2008 funded by FCT/MCTES, Portugal. TLC also acknowledges the support of the UK Science and Technology Facilities Council (STFC). TK is supported by the FWO-Flanders under project O6260 - G.0728.11. Funding for the Stellar Astrophysics Centre is provided by The Danish National Research Foundation (Grant agreement no.: DNRF106). The research is supported by the ASTERISK project (ASTERoseismic Investigations with SONG and Kepler) funded by the European Research Council (Grant agreement no.: 267864).

\clearpage
\begin{table}
\caption{Interpretation of the physical processes responsible for the different features in the solar frequency spectrum used by \citet{1998ESASP.418...83A, 1998ESASP.418...91A, 2004ESASP.538..215A, 2009A&A...495..979M}}
\centering
\begin{tabular}{lc}
\hline \hline
0 -- 1 $\mu$Hz & Rotation\\
10 -- 100 $\mu$Hz & Supergranulation\\
80 -- 1000 $\mu$Hz & Mesogranulation\\
800 -- 3000 $\mu$Hz & Granulation\\
2000 -- 4000 $\mu$Hz & Oscillations\\
\hline
\end{tabular}
\label{tab00}
\end{table}

\begin{table}
\caption{Interpretation of the physical processes responsible for the different features in the solar frequency spectrum used by \citet{1993ASPC...42..111H, 2002ESASP.506..897V, 2012MNRAS.tmp.2467K}}
\centering
\begin{tabular}{lc}
\hline \hline
0 -- 1 $\mu$Hz & Rotation\\
100 -- 1000 $\mu$Hz & Granulation\\
1000 -- 3000 $\mu$Hz & Faculae\\
2000 -- 4000 $\mu$Hz & Oscillations\\
\hline
\end{tabular}
\label{tab00}
\end{table}

\begin{figure}
\epsscale{0.5}
\plotone{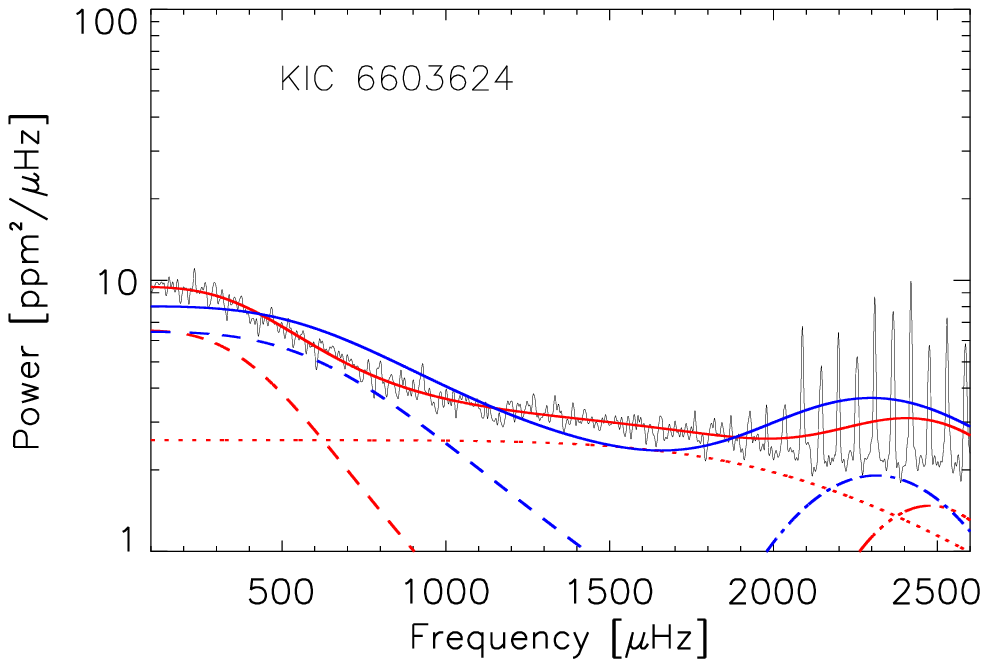}
\plotone{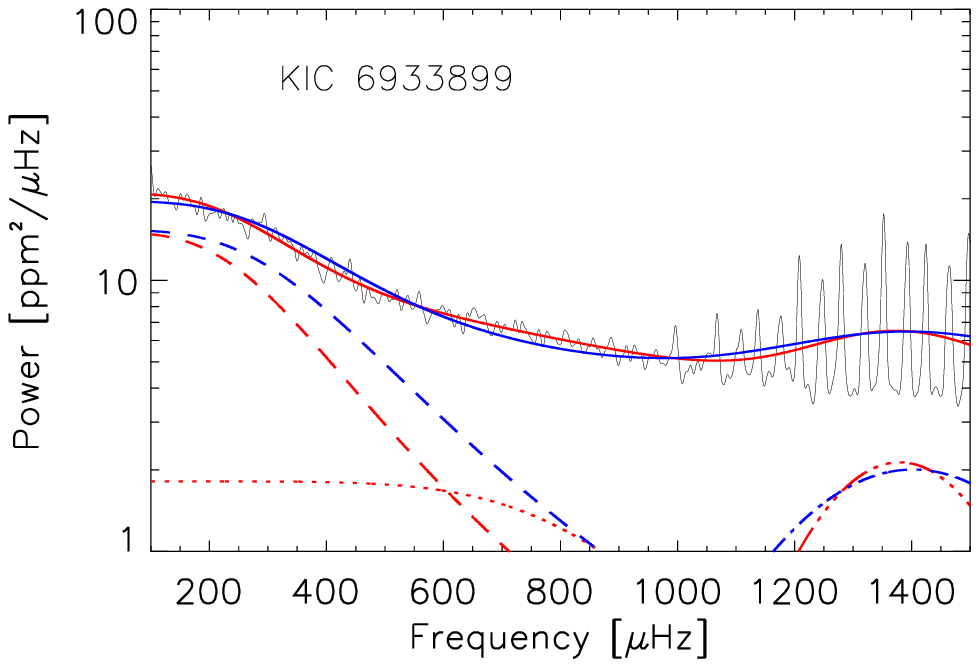}
\plotone{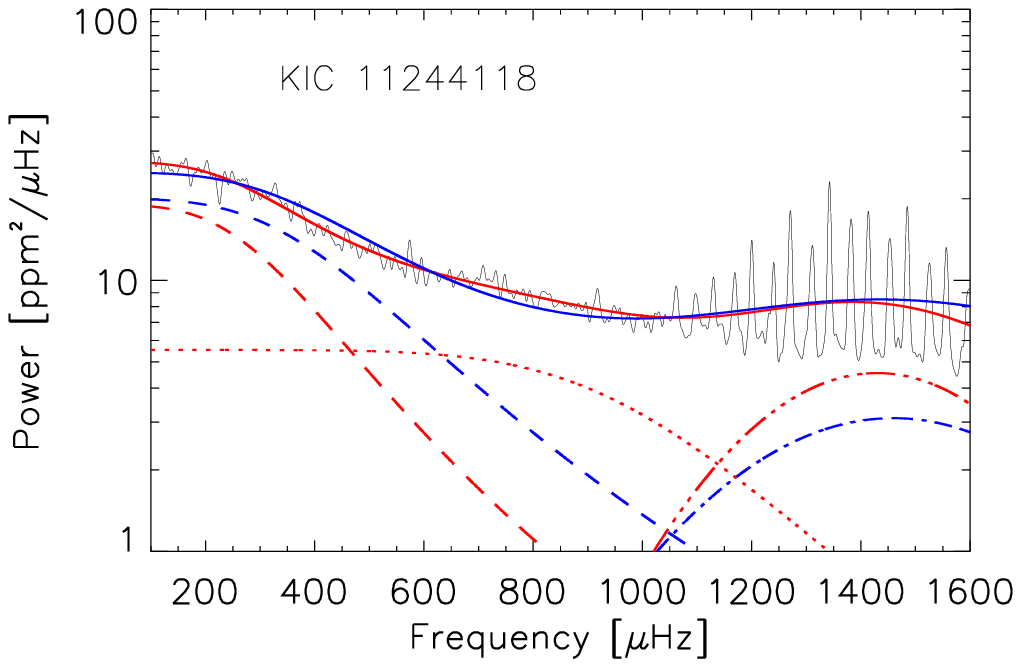}
\caption{Binned average spectra of KIC 6603624, KIC 6933899 and KIC 11244118 (from top to bottom). The solid red lines are the model with the component attributed to faculae and solid blue lines are the model without the component attributed to faculae. The dashed lines show the component attributed to granulation, whereas the dotted lines show the component attributed to faculae and the dash-dotted lines show the p-mode oscillations envelope.}
\end{figure}

\begin{table}
\caption{Stellar parameters from asteroseismology \citep[from][]{2012arXiv1202.2844M, 2011MNRAS.415.3539V}}
\centering
\begin{tabular}{lccc}
\hline \hline
KIC & 6603624 & 6933899 & 11244118 \\
\hline
$R$(R$_{\sun}$) &1.15 $\pm$ 0.01 & 1.58 $\pm$ 0.01 & 1.55 $\pm$ 0.01\\
$M$(M$_{\sun}$) & 1.01 $\pm$ 0.01 & 1.10 $\pm$ 0.01 & 1.01 $\pm$ 0.01\\
$T_{\rm eff}$ (K) & 5416 & 5616 & 5507\\
$K_{\rm p}$ (Mag) & 9.1 & 9.6 & 9.7\\
$Age $(Gyr) & 8.51 $\pm$ 0.23 & 6.28 $\pm$ 0.15 & 8.93 $\pm$ 0.04\\
\hline
\end{tabular}
\label{tab1}
\end{table}

\begin{table}
\caption{Observing times. 'a' means:  KIC 6603624, 'b': KIC 6933899 and 'c': KIC 11244118.}
\centering
\begin{tabular}{llccc}
\hline \hline
Quarter & Start Data & & &  \\
\hline
1 & 13 May -- 17 Jun 2009 & a &  & \\
2.1 & 20 Jun -- 19 Jul 2009 &  & b & c\\
5 & 20 Mar  -- 24 Jun 2010 & a & b & c\\
6 & 24 Jun -- 23 Sep 2010 & a & b & c\\
7 & 23 Sep 2010 -- 6 Jan 2011 & a & b & c\\
8 & 6 Jan 2011 -- 20 Mar 2011 & a & b & c\\
\hline
\end{tabular}
\label{tab1}
\end{table}

\begin{table}
\caption{Logarithmic likelihood ratios [ln($\Lambda$)] -- with fixed and free exponents. }
\centering
\begin{tabular}{lccc}
\hline \hline
KIC & 6603624 & 6933899 & 11244118 \\
\hline
ln($\Lambda)_{\rm fixed}$ &-531.67 & -420.3 & -98.35\\
ln($\Lambda)_{\rm free}$ &-683.16 & -21.34 & -53.34\\ 
\hline
\end{tabular}
\label{tab1}
\end{table}

\begin{table}
{\small
\caption{Mean parameters of the three stars and the Sun. Definitions of the different parameters are given by eq. 2. The number in brackets are the uncertainties returned by the {\it MultiNest} algorithm.}
\centering
\begin{tabular}{lrlrlrlll}
\hline \hline
                                       & KIC 6603624                &            & KIC 6933899                &             & KIC 1124418                   &           & Sun                                 & \\ 
\hline 
 $\sigma_{\rm gran}$ &    62.8 $\pm$      1.5     &(1)       &     74.5 $\pm$      2.3    &(1)        &    87.4 $\pm$      2.9        &(1)      &    62.4 $\pm$      0.6       & ppm  \\
 $\tau_{\rm gran}$      &   280.8 $\pm$      5.9    &(4)       &   474.3 $\pm$     15.4  &(10)      &      448.0 $\pm$     10.7  &(3)      &   214.3 $\pm$     2.9     & sec  \\
 $\sigma_{\rm fac}$    &    76.5 $\pm$      2.4     &(1)        &    28.7 $\pm$      7.5   &(1)         &     74.7 $\pm$        3.1     &(1)      &  50.1 $\pm$      0.1         & ppm  \\
 $\tau_{\rm fac}$         &   66.1 $\pm$      1.3       &(3)       &   180.1 $\pm$      2.4  &(7)         &       153.1 $\pm$      1.6   &(7)      &  65.8 $\pm$      0.3       & sec  \\
 $H_{\rm osc}$            &     1.4 $\pm$      0.1       &(0.03) &     2.1 $\pm$      0.2     &(0.09)   &     4.5 $\pm$           0.3     &(0.1)  &   6.2 $\pm$      2.5         & ppm$^2$/$\mu$Hz  \\
$\nu_{max}$                &  2477 $\pm$     12        &(8)      &  1368 $\pm$      10      &(5)        &           1426 $\pm$    14  &(5)       &  3104 $\pm$      36        & $\mu$Hz  \\
$w$                               &   246.5 $\pm$      0.9    &(10)    &   138.9 $\pm$      2.0   &(8)       &     233.5 $\pm$  2.9         &(7)       &   316 $\pm$      36         & $\mu$Hz  \\
$Noise$                       &     0.4 $\pm$      0.1        & (0.1)         &     4.1 $\pm$      0.1      &   (0.1)         &        2.8 $\pm$      0.4       &    (0.1)       &                                         & ppm$^2$/$\mu$Hz  \\
\hline
\end{tabular}
\label{tab3}
}
\end{table}

\begin{figure}
\epsscale{0.5}
\plotone{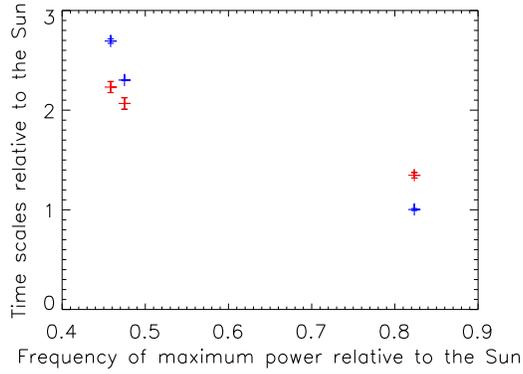}
\caption{Measured time scales of the components attributed to granulation (red) and faculae (blue) as a function of the frequency of maximum power relative to the Sun. It is generally seen that both time scales decreases with increasing frequency of maximum power. }
\end{figure}

\begin{figure}
\epsscale{0.5}
\plotone{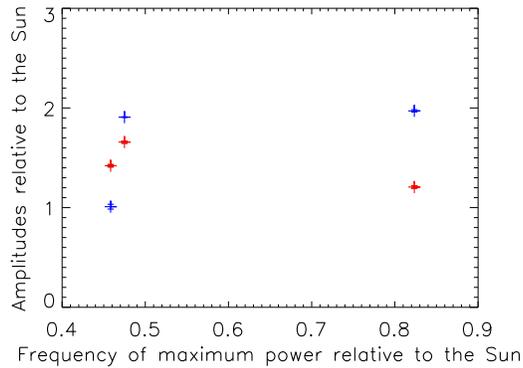}
\caption{Measured amplitudes of the components attributed to granulation (red) and faculae (blue) as a function of the frequency of maximum power relative to the Sun.  No general trend can be seen.}
\end{figure}

\begin{figure}
\epsscale{0.5}
\plotone{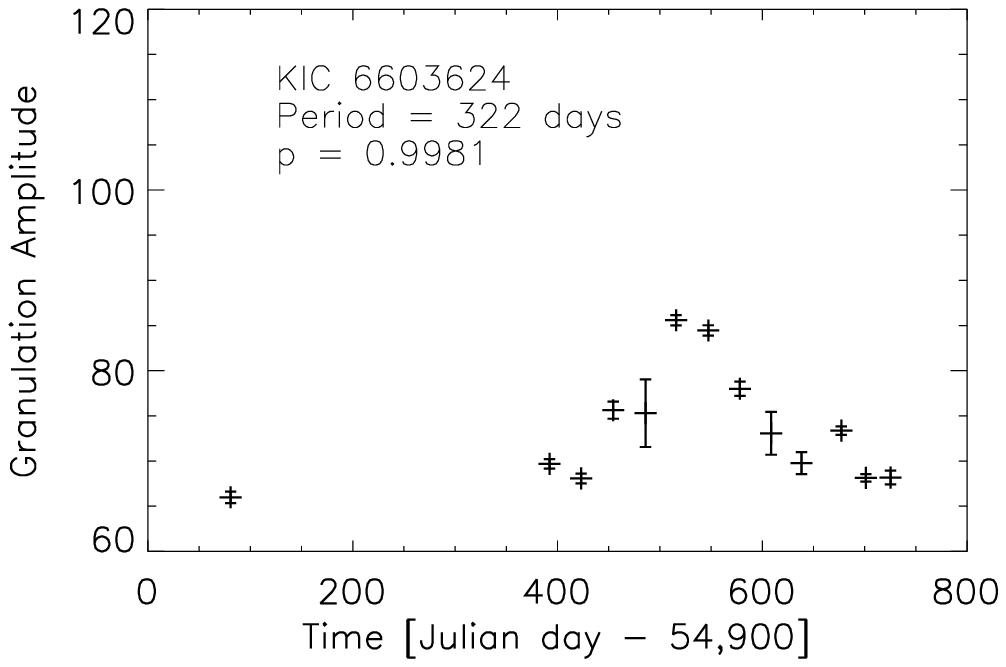}
\plotone{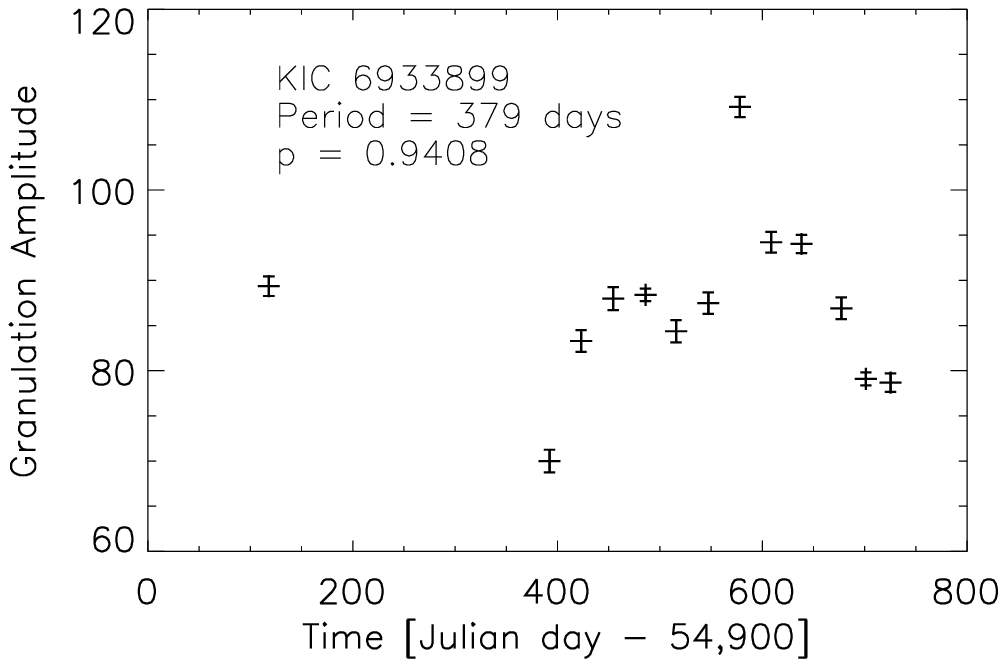}
\plotone{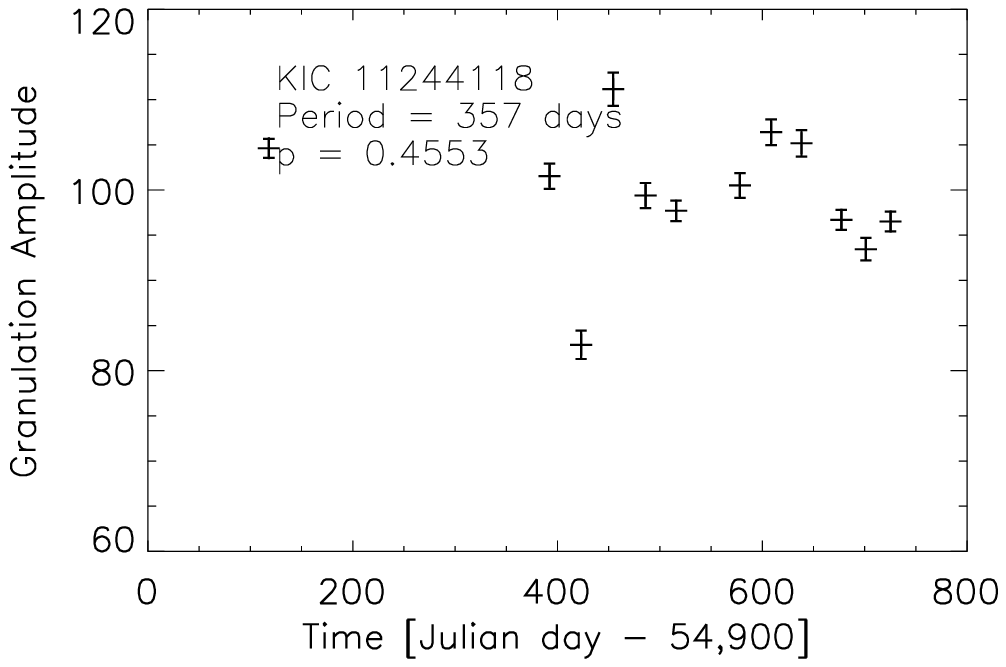}
\caption{Temporal variability of the amplitude of the component attributed to granulation. Periodic variability with a significance level above 0.99 and a period of $322 \pm 11$ days is seen in KIC 6603624.}
\end{figure}

\begin{figure}
\epsscale{0.5}
\plotone{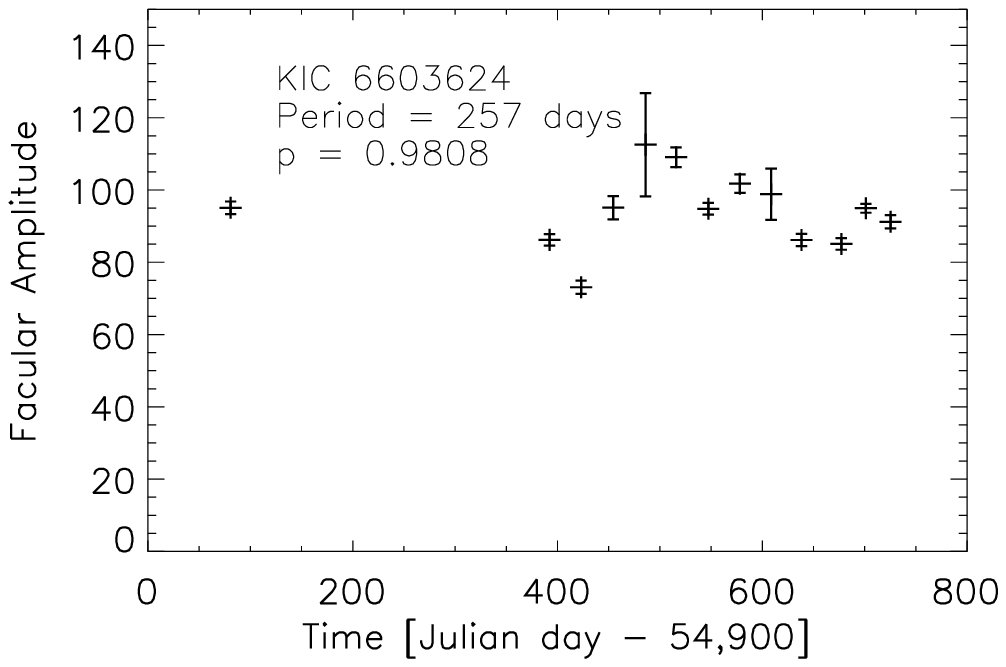}
\plotone{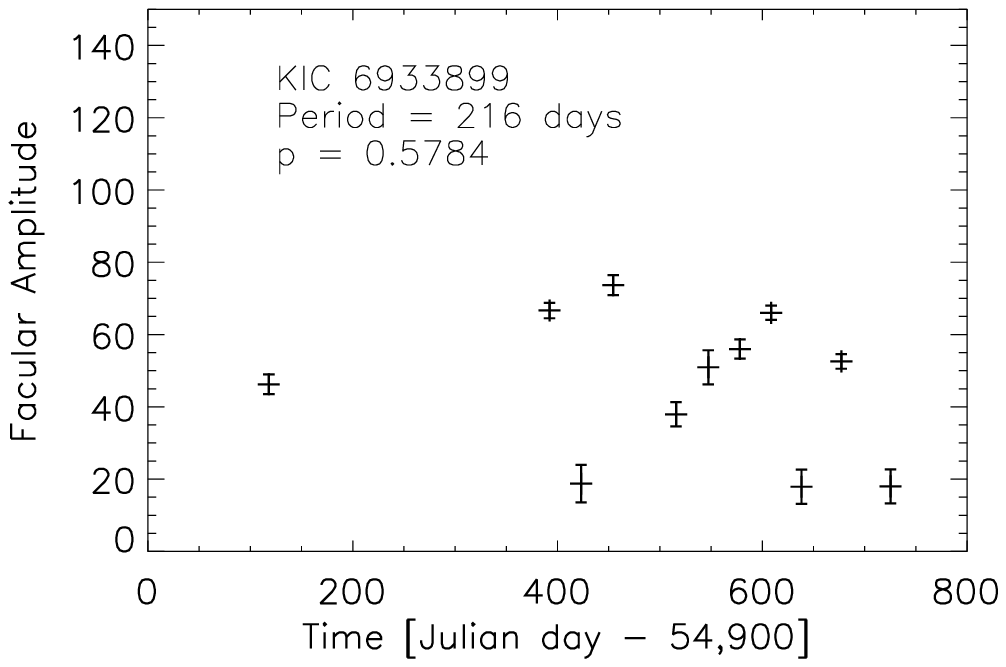}
\plotone{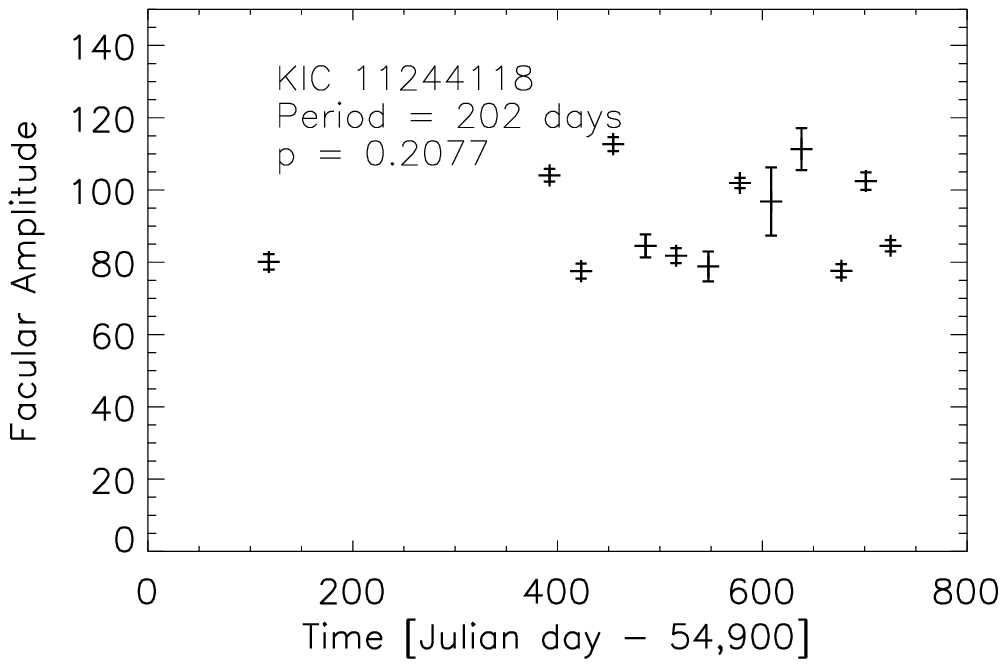}
\caption{Temporal variability of the amplitude of the component attributed to faculae. Marginal significant periodic variability with a significance level above 0.95 and a period of $257 \pm 11$ days is seen in KIC 6603624.}
\end{figure}

\begin{figure}
\epsscale{0.5}
\plotone{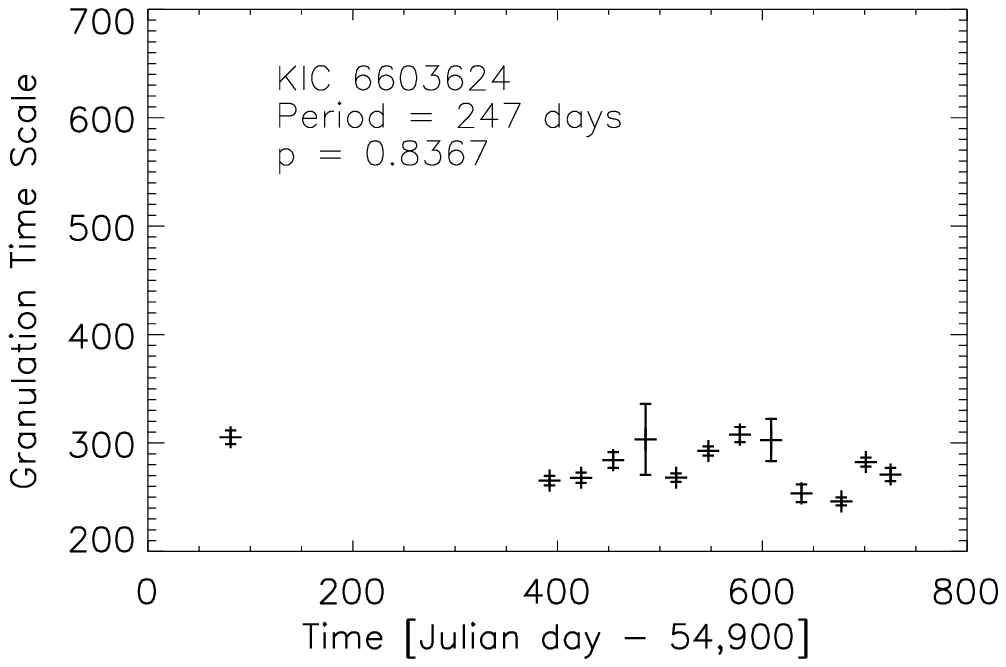}
\plotone{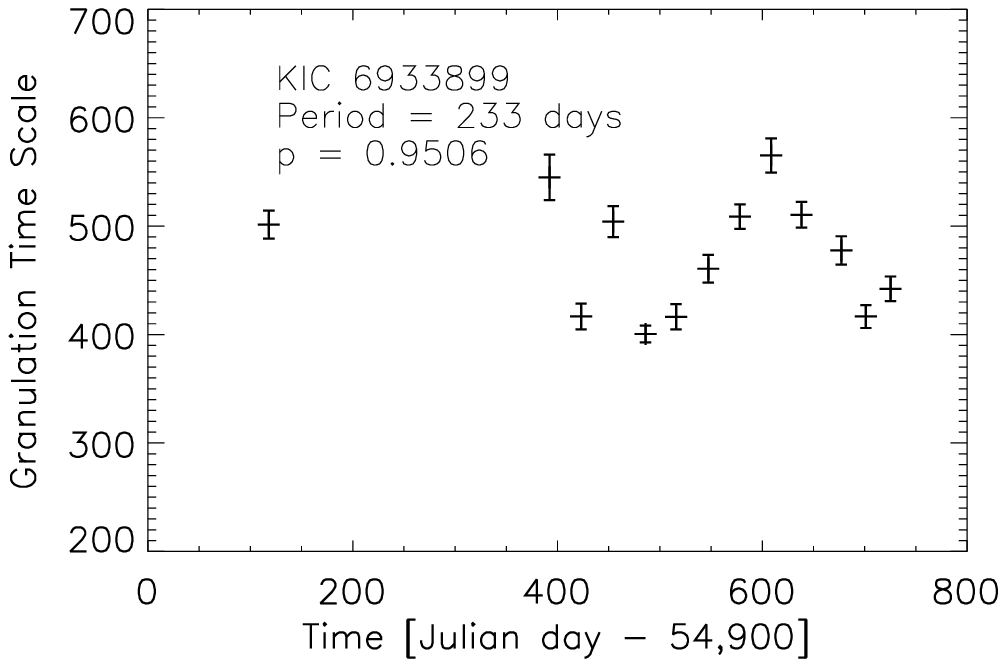}
\plotone{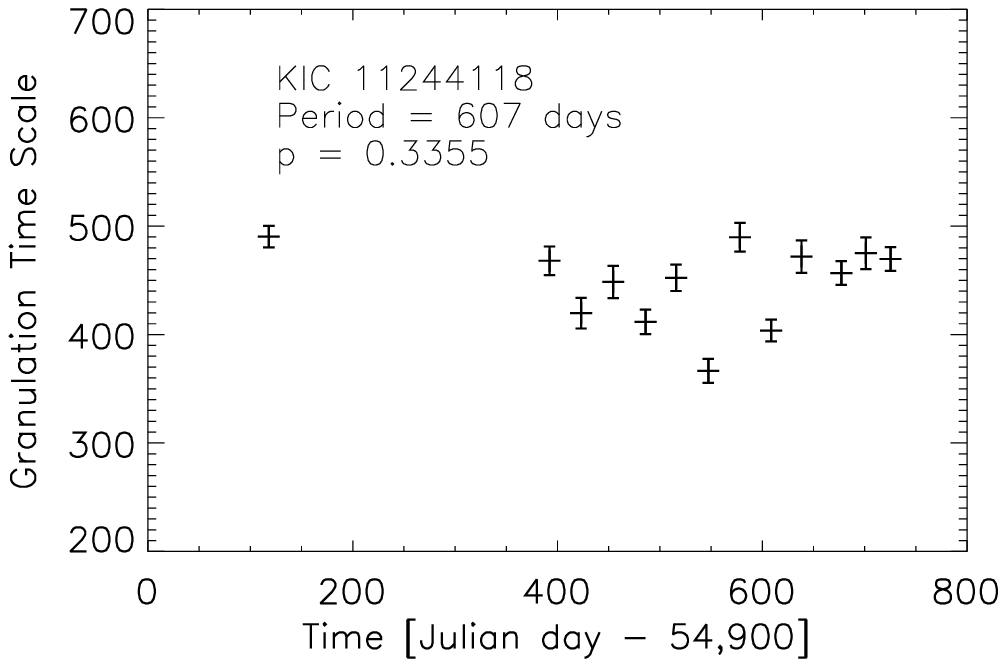}
\caption{Temporal variability of the granulation time scale. Marginal significant periodic variability with a significance level above 0.95 and a period of $233 \pm 13$ days is seen in KIC 6933899.}
\end{figure}

\begin{figure}
\epsscale{0.5}
\plotone{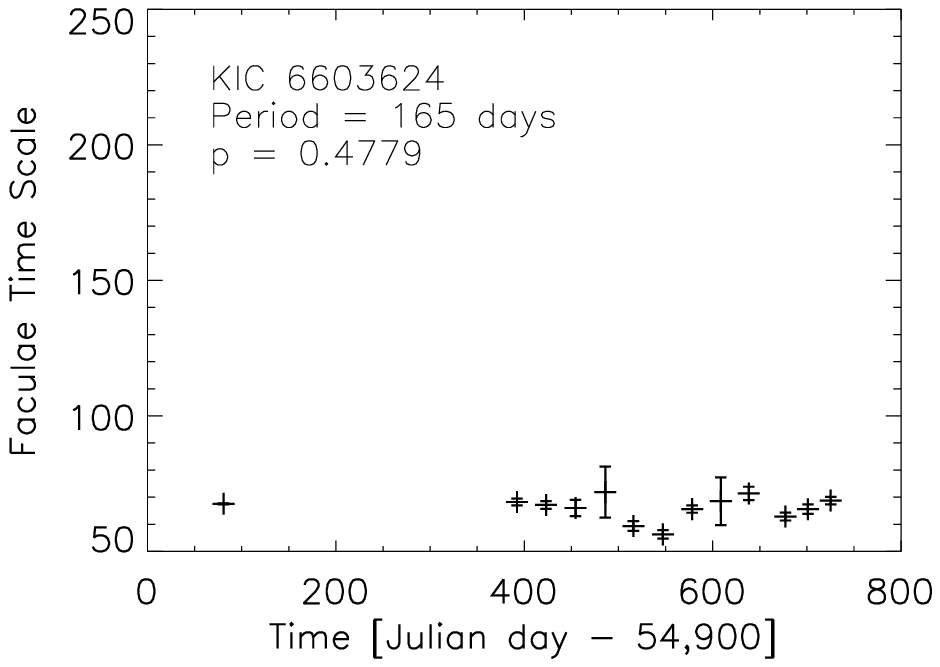}
\plotone{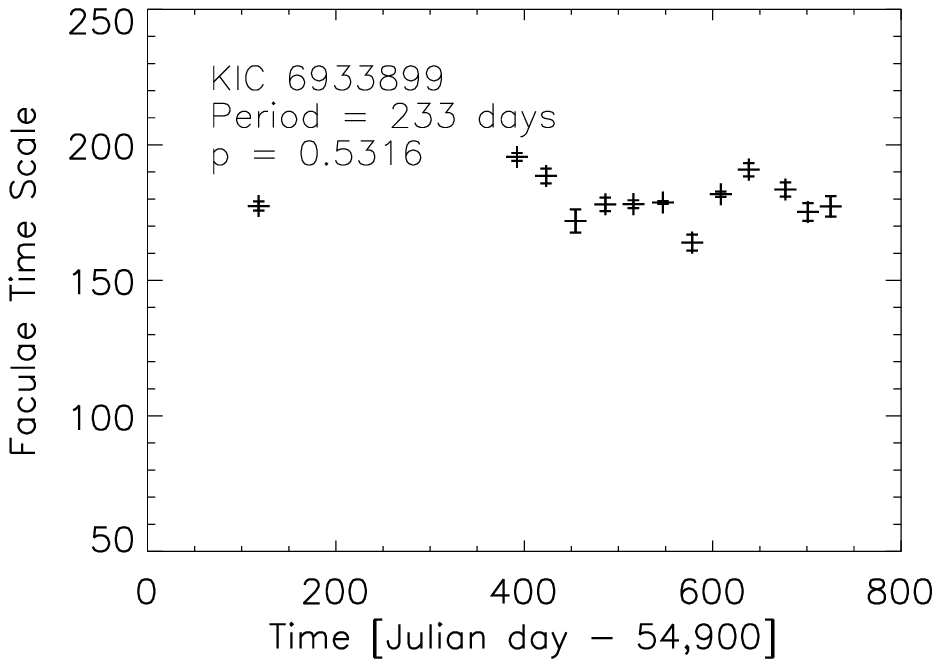}
\plotone{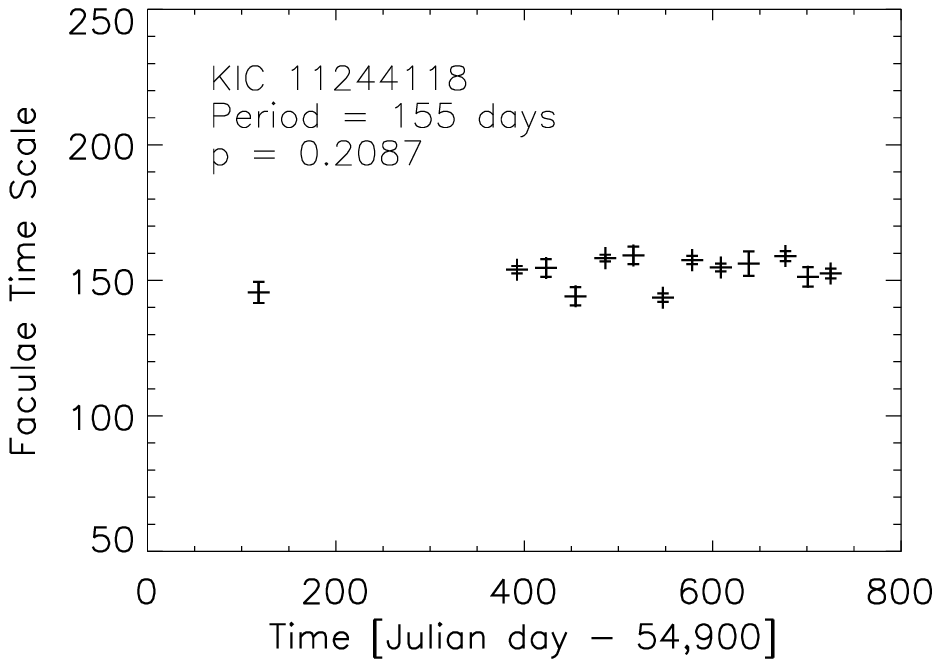}
\caption{Temporal variability of the facular time scale.}
\end{figure}

\begin{figure}
\epsscale{0.5}
\plotone{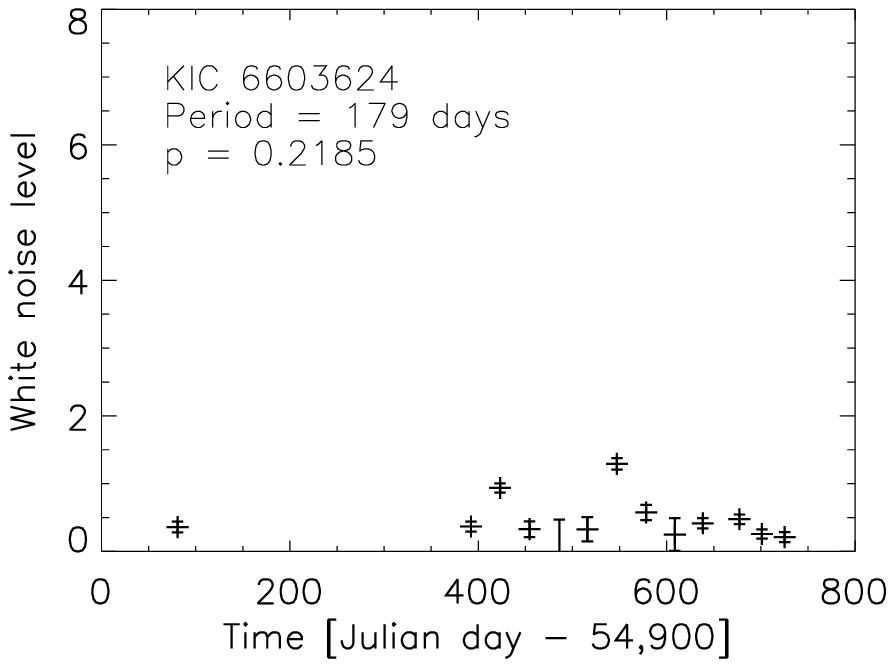}
\plotone{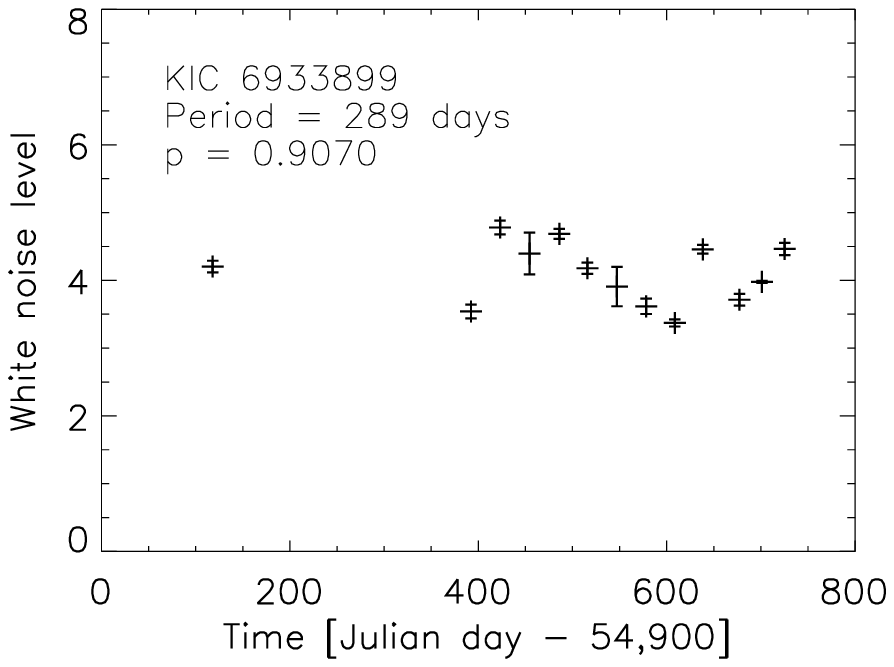}
\plotone{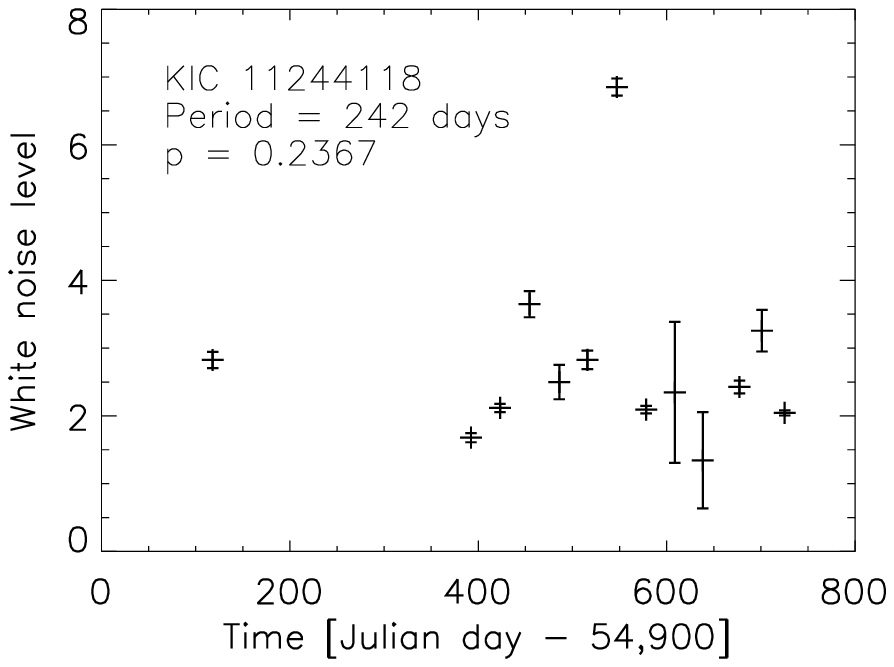}
\caption{Temporal variability of the white noise level.}
\end{figure}

\begin{table}
{\small
\caption{Correlation between the faculae and granulation parameters (KIC 6603624 / KIC 6933899 / KIC 11244118)}
\centering
\begin{tabular}{|l|c|c|c|}
\hline 
					& $\sigma_{\rm fac}$	&  $\tau_{\rm gran}$ 		& $\tau_{\rm fac}$ 		\\
\hline					
$\sigma_{\rm gran}$ 	& 0.56 / 0.21 / 0.09	& 0.15 / 0.26 / -0.47		& -0.72 / -0.56 / -0.61	\\ 
\hline
$\sigma_{\rm fac}$		&				& 0.59 / 0.74 / 0.36		& -0.09 / 0.01 / -0.08		\\
\hline
$\tau_{\rm gran}$		&				&					& 0.06 / 0.18 / 0.18  		\\
\hline
\end{tabular}
\label{tab3}
}
\end{table}

\clearpage


\begin{thebibliography}{}
\bibitem[Aigrain et al.(2004)]{2004ESASP.538..215A} Aigrain, S., Favata, F., \& Gilmore, G.\ 2004, Stellar Structure and Habitable Planet Finding, 538, 215 
\bibitem[Andersen et al.(1998)]{1998ESASP.418...83A} Andersen, B., Leifsen, T., Appourchaux, T., et al.\ 1998, Structure and Dynamics of the Interior 
of the Sun and Sun-like Stars, 418, 83 
\bibitem[Anklin et al.(1998)]{1998ESASP.418...91A} Anklin, M., Frohlich, C., Wehrli, C., \& Finsterle, W.\ 1998, Structure and Dynamics of the Interior of the Sun and Sun-like Stars, 418, 91
\bibitem[Antalova(1994)]{1994AdSpR..14..721A} Antalova, A.\ 1994, Advances in Space Research, 14, 721 
\bibitem[Appourchaux(2003)]{2003A&A...412..903A} Appourchaux, T.\ 2003, \aap, 412, 903 
\bibitem[Appourchaux(2004)]{2004A&A...428.1039A} Appourchaux, T.\ 2004, A\&A, 428, 1039 
\bibitem[Appourchaux et al.(2012)]{2012arXiv1204.3147A} Appourchaux, T., Chaplin, W.~J., Garc{\'{\i}}a, R.~A., et al.\ 2012, arXiv:1204.3147 
\bibitem[Belkacem et al.(2011)]{2011A&A...530A.142B} Belkacem, K., Goupil, M.~J., Dupret, M.~A., et al.\ 2011, A\&A, 530, A142 
\bibitem[Broomhall et al.(2012)]{2012MNRAS.420.1405B} Broomhall, A.-M., Chaplin, W.~J., Elsworth, Y., \& Simoniello, R.\ 2012, \mnras, 420, 1405 
\bibitem[Brown et al.(1991)]{1991ApJ...368..599B} Brown, T.~M., Gilliland, R.~L., Noyes, R.~W., \& Ramsey, L.~W.\ 1991, \apj, 368, 599 
\bibitem[Bruntt et al.(2012)]{2012arXiv1203.0611B} Bruntt, H., Basu, S., Smalley, B., et al.\ 2012, arXiv:1203.0611 
\bibitem[Chaplin et al.(2010)]{2010ApJ...713L.169C} Chaplin, W.~J., et al.\ 2010, ApJ, 713, L169
\bibitem[Chaplin et al.(2011a)]{2011Sci...332..213C} Chaplin, W.~J., Kjeldsen, H., Christensen-Dalsgaard, J., et al.\ 2011, Science, 332, 213 
\bibitem[Chaplin et al.(2011b)]{2011ApJ...732...54C} Chaplin, W.~J., Kjeldsen, H., Bedding, T.~R., et al.\ 2011, \apj, 732, 54 
\bibitem[Del Moro(2004)]{2004A&A...428.1007D} Del Moro, D.\ 2004, A\&A, 428, 1007 
\bibitem[Fletcher et al.(2010)]{2010ApJ...718L..19F} Fletcher, S.~T., Broomhall, A.-M., Salabert, D., Basu, S., Chaplin, W.~J., Elsworth, Y., Garc{\'{\i}}a, R.~A., \& New, R.\ 2010, ApJ, 718, L19 
\bibitem[Fr{\"o}hlich et al.(1995)]{1995SoPh..162..101F} Fr{\"o}hlich, C., Romero, J., Roth, H., et al.\ 1995, \solphys, 162, 101 
\bibitem[Gabriel(1994)]{1994A&A...287..685G} Gabriel, M.\ 1994, \aap, 287, 685 
\bibitem[Gabriel et al.(1995)]{1995SoPh..162...61G} Gabriel, A.~H., Grec, G., Charra, J., et al.\ 1995, \solphys, 162, 61 
\bibitem[Garc{\'{\i}}a et al.(1998)]{1998ApJ...504L..51G} Garc{\'{\i}}a, R.~A., Pall\'e, P.~L., Turck-Chi\`eze, S., et al.\ 1998, \apjl, 504, L51 
\bibitem[Garc{\'{\i}}a et al.(2011)]{2011MNRAS.414L...6G} Garc{\'{\i}}a, R.~A., Hekker, S., Stello, D., et al.\ 2011, \mnras, 414, L6 
\bibitem[Gilliland et al.(2010)]{2010ApJ...713L.160G} Gilliland, R.~L., Jenkins, J.~M., Borucki, W.~J., et al.\ 2010, \apjl, 713, L160 
\bibitem[Goldreich et al.(1994)]{1994ApJ...424..466G} Goldreich, P., Murray, N., \& Kumar, P.\ 1994, \apj, 424, 466 
\bibitem[Harvey(1985)]{1985shpp.rept..199H} Harvey, J.\ 1985, Future Missions in Solar, Heliospheric and Space Plasma Physicenvelopes, 235, 199 
\bibitem[Harvey et al.(1993)]{1993ASPC...42..111H} Harvey, J.~W., Duvall, T.~L., Jr., Jefferies, S.~M., \& Pomerantz, M.~A.\ 1993, GONG 1992.~Seismic Investigation of the Sun and Stars, 42, 111
\bibitem[Houdek et al.(1999)]{1999A&A...351..582H} Houdek, G., Balmforth, N.~J., Christensen-Dalsgaard, J., \& Gough, D.~O.\ 1999, \aap, 351, 582
\bibitem[Howe et al.(2000)]{2000Sci...287.2456H} Howe, R., Christensen-Dalsgaard, J., Hill, F., et al.\ 2000, Science, 287, 2456 
\bibitem[Jenkins et al.(2010)]{2010ApJ...713L..87J} Jenkins, J.~M., Caldwell, D.~A., Chandrasekaran, H., et al.\ 2010, \apjl, 713, L87 
\bibitem[Jim{\'e}nez et al.(2005)]{2005ApJ...623.1215J} Jim{\'e}nez, A., Jim{\'e}nez-Reyes, S.~J., \& Garc{\'{\i}}a, R.~A.\ 2005, \apj, 623, 1215 
\bibitem[Kjeldsen \& Bedding(1995)]{1995A&A...293...87K} Kjeldsen, H., \& Bedding, T.~R.\ 1995, \aap, 293, 87 
\bibitem[Kjeldsen \& Bedding(2011)]{2011A&A...529L...8K} Kjeldsen, H., \& Bedding, T.~R.\ 2011, \aap, 529, L8 
\bibitem[Kallinger et al.(2010)]{2010A&A...522A...1K} Kallinger, T., Mosser, B., Hekker, S., et al.\ 2010, \aap, 522, A1 
\bibitem[Karoff(2007)]{2007MNRAS.381.1001K} Karoff, C.\ 2007, \mnras, 381, 1001 
\bibitem[Karoff(2008)]{karoff2008} Karoff, C.\ 2008, PhD dissertation, Aarhus University
\bibitem[Karoff(2009)]{2009ASPC..416..233K} Karoff, C.\ 2009, Solar-Stellar Dynamos as Revealed by Helio- and Asteroseismology: GONG 2008/SOHO 21, 416, 233 
\bibitem[Karoff(2012)]{2012MNRAS.tmp.2467K} Karoff, C.\ 2012, \mnras, 421, 3170 
\bibitem[Karoff \& Kjeldsen(2008)]{2008ApJ...678L..73K} Karoff, C., \& Kjeldsen, H.\ 2008, \apjl, 678, L73
\bibitem[Koch et al.(2010)]{2010ApJ...713L..79K} Koch, D.~G., Borucki, W.~J., Basri, G., et al.\ 2010, \apjl, 713, L79 
\bibitem[Lanza et al.(2004)]{2004A&A...425..707L} Lanza, A.~F., Rodon{\`o}, M., \& Pagano, I.\ 2004, \aap, 425, 707 
\bibitem[Lefebvre et al.(2008)]{2008A&A...490.1143L} Lefebvre, S., Garc{\'{\i}}a, R.~A., Jim{\'e}nez-Reyes, S.~J., Turck-Chi{\`e}ze, S., \& Mathur, S.\ 2008, A\&A, 490, 1143
\bibitem[Lomb(1976)]{1976Ap&SS..39..447L} Lomb, N.~R.\ 1976, Ap\&SS, 39, 447
\bibitem[Mathur et al.(2010)]{2010A&A...511A..46M} Mathur, S., Garc{\'{\i}}a, R.~A., R{\'e}gulo, C., et al.\ 2010, \aap, 511, A46
\bibitem[Mathur et al.(2011)]{2011ApJ...741..119M} Mathur, S., Hekker, S., Trampedach, R., et al.\ 2011, \apj, 741, 119
\bibitem[Mathur et al.(2012)]{2012arXiv1202.2844M} Mathur, S., Metcalfe, T.~S., Woitaszek, M., et al.\ 2012, \apj, 749, 152 
\bibitem[Metcalfe et al.(2009)]{2009ApJ...699..373M} Metcalfe, T.~S., Creevey, O.~L., \& Christensen-Dalsgaard, J.\ 2009, \apj, 699, 373 
\bibitem[Michel et al.(2008)]{2008Sci...322..558M} Michel, E., Baglin, A., Auvergne, M., et al.\ 2008, Science, 322, 558
\bibitem[Michel et al.(2009)]{2009A&A...495..979M} Michel, E., Samadi, R., Baudin, F., et al.\ 2009, \aap, 495, 979 
\bibitem[Mursula et al.(2003)]{2003SoPh..212..201M} Mursula, K., Zieger, B., \& Vilppola, J.~H.\ 2003, \solphys, 212, 201 
\bibitem[Nordlund et al.(1997)]{1997A&A...328..229N} Nordlund, A., Spruit, H.~C., Ludwig, H.-G., \& Trampedach, R.\ 1997, A\&A, 328, 229
\bibitem[Press et al.(1992)]{1992nrfa.book.....P} Press, W.~H., Teukolsky, S.~A., Vetterling, W.~T., \& Flannery, B.~P.\ , Numerical recipes in C, Cambridge: University Press, 1992, 2nd ed., 
\bibitem[Richardson et al.(1994)]{1994GeoRL..21.1559R} Richardson, J.~D., Paularena, K.~I., Belcher, J.~W., \& Lazarus, A.~J.\ 1994, \grl, 21, 1559 
\bibitem[Roudier et al.(1998)]{1998A&A...330.1136R} Roudier, T., Malherbe, J.~M., Vigneau, J., \& Pfeiffer, B.\ 1998, A\&A, 330, 1136 
\bibitem[Sakurai(1979)]{1979Natur.278..146S} Sakurai, K.\ 1979, \nat, 278, 146 
\bibitem[Schwarzschild(1975)]{1975ApJ...195..137S} Schwarzschild, M.\ 1975, \apj, 195, 137 
\bibitem[Simoniello et al.(2012)]{2012arXiv1210.6796S} Simoniello, R., Jain, K., Tripathy, S.~C., et al.\ 2012, arXiv:1210.6796 
\bibitem[Skumanich(1972)]{1972ApJ...171..565S} Skumanich, A.\ 1972, \apj, 171, 565
\bibitem[Stello et al.(2009)]{2009MNRAS.400L..80S} Stello, D., Chaplin, W.~J., Basu, S., Elsworth, Y., \& Bedding, T.~R.\ 2009, \mnras, 400, L80 
\bibitem[Vald{\'e}s-Galicia et al.(1996)]{1996SoPh..167..409V} Vald{\'e}s-Galicia, J.~F., P{\'e}rez-Enr{\'{\i}}quez, R., \& Otaola, J.~A.\ 1996, \solphys, 167, 409 
\bibitem[Vald{\'e}s-Galicia \& Velasco(2008)]{2008AdSpR..41..297V} Vald{\'e}s-Galicia, J.~F., \& Velasco, V.~M.\ 2008, Advances in Space Research, 41, 297
\bibitem[V{\'a}zquez Rami{\'o} et al.(2002)]{2002ESASP.506..897V} V{\'a}zquez Rami{\'o}, H., Roca Cort{\'e}s, T., \& R{\'e}gulo, C.\ 2002, Solar Variability: From Core to Outer Frontiers, 506, 897 
\bibitem[Verner et al.(2011)]{2011MNRAS.415.3539V} Verner, G.~A., Elsworth, Y., Chaplin, W.~J., et al.\ 2011, \mnras, 415, 3539 
\bibitem[Quirion et al.(2010)]{2010ApJ...725.2176Q} Quirion, P.-O., Christensen-Dalsgaard, J., \& Arentoft, T.\ 2010, \apj, 725, 2176 
\end{thebibliography}
\end{document}